\documentclass[RNAAS]{aastex}

\usepackage{natbib}
\begin{document}

\title{Opportunities for the Large Synoptic Survey Telescope  to Find New L$_5$ Trojan and Hilda \emph{Lucy} Encounter Targets}

\author[0000-0003-4365-1455]{Megan E. Schwamb}
\affiliation{Gemini Observatory,  Northern Operations Center, 670 North A'ohoku Place, Hilo, HI 96720, USA}

\author{Harold F. Levison}
\affiliation{Southwest Research Institute 1050 Walnut Street, Suite 300 Boulder, CO 80302, USA}

\author[0000-0003-0854-745X]{Marc W. Buie}
\affiliation{Southwest Research Institute 1050 Walnut Street, Suite 300 Boulder, CO 80302, USA}

\correspondingauthor{Megan E. Schwamb}
\email{mschwamb.astro@gmail.com}
\section*{}

\emph{Lucy} \citep{2016LPI....47.2061L} is an upcoming NASA Discovery
mission to explore the Jupiter Trojans, asteroids locked in the 1:1 mean motion resonance with Jupiter
\citep{2002aste.book..273B,2002aste.book..725M,2004jpsm.book..263J}.
In the next two decades, Lucy will explore several Jupiter Trojans in the leading (L$_4$) and trailing (L$_5$) clouds --- Lagrangian regions of gravitational equilibrium within Jupiter's orbit flanking ahead and behind the planet by 60 degrees. The \emph{Lucy} primary mission timeline is as follows:
\begin{itemize}
\item \emph{Lucy} will launch in October or November of 2021.
\item In April 2025, \emph{Lucy} will encounter Main Belt asteroid (52246) Donaldjohanson on the way to the L$_4$ cloud.
\item From 2027-2028, the spacecraft will fly by four L$_4$ Trojans: (3548) Eurybates, (15094) Polymele, (11351) Leucus, and (21900) Orus. 
\item In 2033, \emph{Lucy} will encounter the (617) Patroclus-Menoetius (PM) Trojan binary in the L$_5$ cloud. 
\end{itemize}
The mission will probe the cratering history, geology, surface composition, masses, bulk densities, internal structure, satellite and ring fraction, and other characteristics of these primitive small bodies (see \url{http://lucy.swri.edu} for more information). 

The Trojan clouds contain a surprisingly diverse population of objects.  Only by understanding this diversity can we unravel their history.  The more objects \emph{Lucy} visits the better we can  constrain the formation and evolution of the giant planets.  While \emph{Lucy} will visit four Trojans in the L$_4$ cloud, its trajectory only passes one known Trojan system, the PM binary, in the L$_5$ cloud.  There is ample time in the \emph{Lucy} timeline to possibly add additional L$_5$ targets, thereby maximizing the mission's science return.  In the next decade, the 8.4-m Large Synoptic Survey Telescope \citep[LSST;][]{2008arXiv0805.2366I}, is expected to detect nearly 300,000 Jupiter Trojans between approximately 16 and 24.5 r-band magnitudes in the $\sim$18,000 deg$^2$ main survey  and the expected $\sim$4,000 deg$^2$ northern ecliptic extension \citep{2009arXiv0912.0201L,2017arXiv170804058L}. We explore potential opportunities for LSST to find new \emph{Lucy} L$_5$  fly-by targets.
 
\begin{figure}
\begin{center}
\includegraphics[width=0.9\columnwidth]{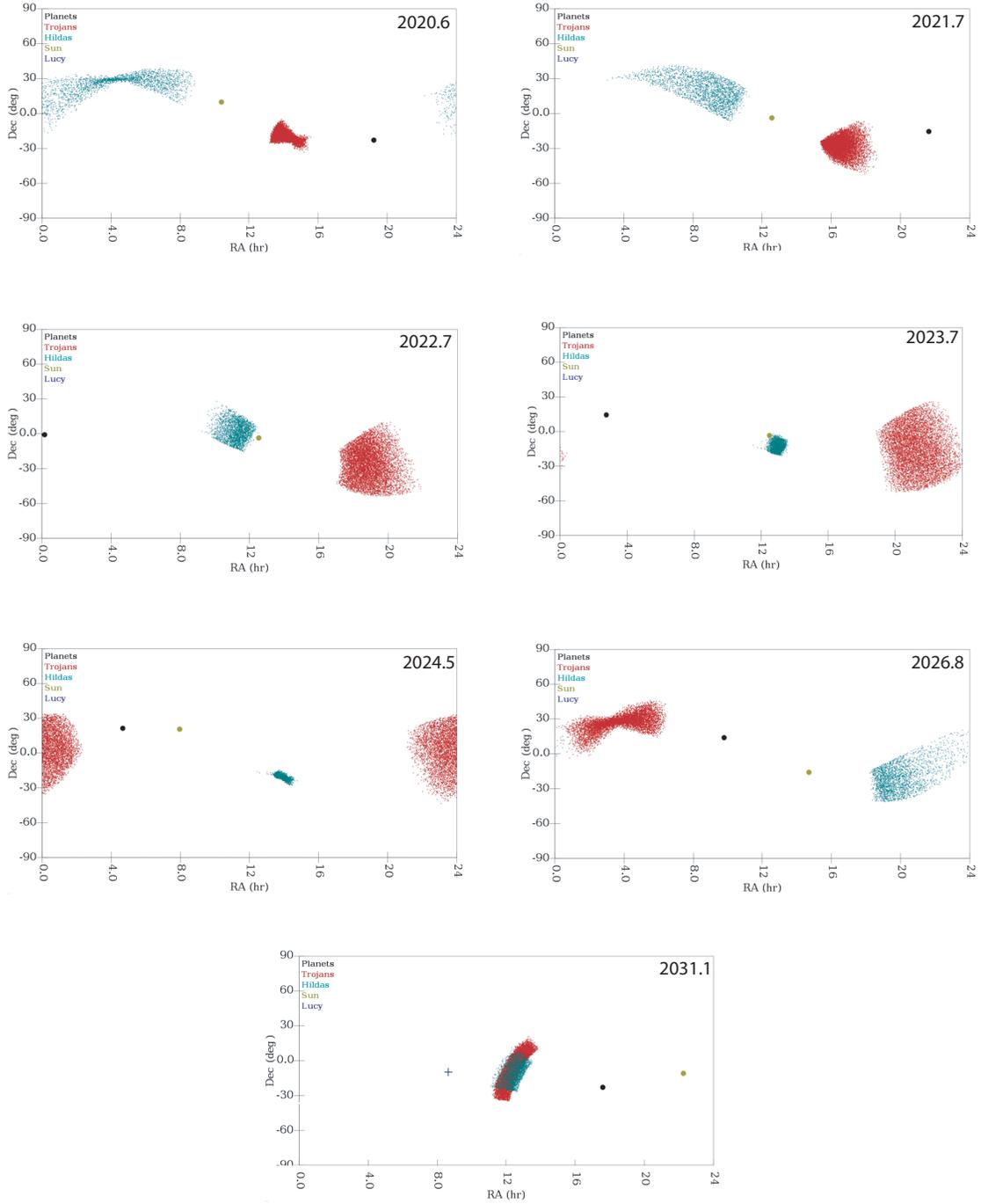}

\caption{\textbf{\label{orbit_snapshots}} A dynamical integration of the 
collection of hypothetical Trojan and Hilda orbits that Lucy would 
encounter in the $L_5$ cloud but before the Patroclus-Menoetius fly-by. 
The static figure shows on-sky positions of the Hildas and Trojans at a 
sampling of LSST operation stages when either the Hildas or Trojans are 
observable. The animated figure (available in the RNAAS published online version) displays the simulation, including the 
distribution looking down on the Solar System (animation, top panel) and 
the location on-sky (animation, bottom panel), running backward from 
year 2033.3 to 2016.9.}
\end{center}
\end{figure}

LSST engineering first light with ComCam, the $\sim$0.7 deg$^2$ field-of-view (FOV) telescope commissioning camera, is expected in 2020 \citep{2018SPIE10700E..3DH}. On-sky commissioning and science verification of the 3.2 Giga-pixel LSST science camera, with a 9.6 deg$^2$ FOV,  is scheduled to begin in 2021 \citep{2018SPIE10705E..0EB}.  Full LSST science operations are planned to start in 2022 with a baseline survey lasting ten years \citep{2008arXiv0805.2366I}. Results from a dynamical integration of orbits that would pass close to \emph{Lucy}  in the L$_5$ cloud are shown in Figure~\ref{orbit_snapshots}. We use a hypothetical population of L$_5$ Trojans, as well as simulated orbits for Hilda asteroids (Hildas; asteroids locked in the 3:2 mean motion resonance with Jupiter near 3.9 au) intersecting the L$_5$  cloud. These simulations were carried out with the  SWIFT RMVS (Regularized Mixed Variable Symplectic) N -body integrator \citep{1994Icar..108...18L}. There are particularly favorable opportunities when the simulated L$_5$ Trojan and Hilda particles collapse to regions covering approximately a few hundred square degrees on the sky.  Many of these are observable from Cerro Pach\'{o}n. Equipped with a 9.6 deg$^2$  FOV camera, LSST can efficiently cover these regions with a tractable number of pointings. This may allow the possibly for repeated imaging that can be stacked and shifted to search for objects fainter than the main LSST survey depth. The potential L$_5$ Trojan \emph{Lucy} encounter orbits will be the most compact in 2020 during LSST commissioning. The same occurs for the Hildas in 2024 during the main survey. The Solar System community and LSST project should consider this unique opportunity when planning commissioning observations, the LSST survey cadence, and field prioritization.

\software{SWIFT \citep{1994Icar..108...18L}}

\acknowledgments

MES was supported by Gemini Observatory which is operated by the Association of Universities for Research in Astronomy, Inc., on behalf of the international Gemini partnership of Argentina, Brazil, Canada, Chile, and the United States of America. We thank the LSST Solar System Science Collaboration and  Chris Lintott for manuscript feedback.

\bibliographystyle{aasjournal}

\end{document}